\documentclass[prd,onecolumn,english]{revtex4}
\usepackage{babel}

\usepackage{amsmath}
\usepackage{amssymb}
\usepackage{amsthm} % ??? ?????????? ??????, ????? ? ?????? ?????????.

\usepackage{array}

\usepackage{slashed}

\usepackage[dvips]{graphicx} % ????????? ??????? ??????????? ??????.
\usepackage{subfigure}

\usepackage{varioref}   % ??????????? ?????? ?? ??????????.
\usepackage{xr}         % ?????? ?? ??????? ?????????.

\usepackage{pstricks}           % PSTricks with the `color' interface

\allowdisplaybreaks[1]

\newcommand\nn{\nonumber}
\newcommand\ba{\begin{eqnarray}}
\newcommand\ea{\end{eqnarray}}
\newcommand\eq[1] {\begin{align} #1 \end{align}}   % ????????? ??? ???????.
\newcommand\ga[1] {\begin{gather} #1 \end{gather}}   % ????????? ??? ???????.

\begin{document}
% =========================================================================

\title{Imaginary part of the zero angle light-by-light scattering tensor}

\author{A.~I.~Ahmadov$^{1,2}$~\footnote{E-mail: ahmadov@theor.jinr.ru}, S.~Bakmaev$^{3}$~\footnote{E-mail: bakmaev@jinr.ru} and E.~A.~Kuraev$^{1}$~\footnote{E-mail: kuraev@theor.jinr.ru}}
\affiliation{$^{1}$ Bogoliubov Laboratory of Theoretical Physics,
Joint Institute for Nuclear Research, Dubna,  141980 Russia \\
$^{2}$ Institute of Physics, Azerbaijan
National Academy of Sciences, Baku, Azerbaijan  \\
$^{3}$Dzhelepov Laboratory of Nuclear Problems,
Joint Institute for Nuclear Research, Dubna,  141980 Russia}

\date{\today}

\begin{abstract}
We calculate the bilinear combination of Dirac tensors describing
the creation of a pair of
charged particle (pi-meson or leptons) by virtual photons integrated
on the final particle phase volume. It can be interpreted as an $s$
channel discontinuity of the zero angle light-by-light scattering
tensor with both photons off mass shell.
The expression of light-by-light scattering tensor is represented in
the form where the gauge invariance and the Bose symmetry are explicitly
shown. Some applications and the checks are discussed.
\end{abstract}

\maketitle

% =========================================================================
\section{Introduction}
\label{SectionIntroduction}
% =========================================================================
A lot of attention was paid to the problem of describing the lowest order
nonlinear effect of Quantum Electrodynamics, the scattering of photon on photon.
One of earlier papers \cite{KN1950} reveals rather a complicated structure of tensor
describing the light-by-light scattering ($lbl$) process
\ba
\gamma^*(k_1,\mu)+\gamma^*(k_2,\nu)\to \gamma^*(k_3,\rho)+\gamma^*(k_4,\sigma)
\ea
with all photons off mass shell have rather cumbersome form:
\ba
G_{\mu\nu\rho\sigma}(1234) = \sum_{24 perm}\{\frac{1}{8}A^{2143}(1234)k_{\mu}^{(2)}
k_{\nu}^{(1)}k_{\rho}^{(4)}k_{\sigma}^{(3)} +\frac{1}{4}A^{2341}(1234)k_{\mu}^{(2)}
k_{\nu}^{(3)}k_{\rho}^{(4)}k_{\sigma}^{(1)} + \nn \\
\frac{1}{2}A^{2111}(1234)k_{\mu}^{(2)}k_{\nu}^{(1)}k_{\rho}^{(1)}k_{\sigma}^{(1)} +
\frac{1}{2}A^{2121}(1234)k_{\mu}^{(2)}k_{\nu}^{(1)}k_{\rho}^{(2)}k_{\sigma}^{(1)} + \nn \\
A^{2311}(1234)k_{\mu}^{(2)}k_{\nu}^{(3)}k_{\rho}^{(1)}k_{\sigma}^{(1)} + A^{2123}(1234)
k_{\mu}^{(2)}k_{\nu}^{(1)}k_{\rho}^{(2)}k_{\sigma}^{(3)} +  \nn \\
\frac{1}{2}B_1^{11}(1234)g_{\mu \nu}k_{\rho}^{(1)}k_{\sigma}^{(1)} +
\frac{1}{2}B_1^{12}(1234)g_{\mu \nu}k_{\rho}^{(1)}k_{\sigma}^{(2)} +  \nn \\
B_1^{13}(1234)g_{\mu \nu}k_{\rho}^{(1)}k_{\sigma}^{(3)} +
\frac{1}{4}B_1^{43}(1234)g_{\mu \nu}k_{\rho}^{(4)}k_{\sigma}^{(3)} +
\frac{1}{8}C_1(1234)g_{\mu \nu} g_{\rho \sigma}\},
\ea
with the notation given in \cite{KN1950}.
Detailed investigations of the light-by-light scattering tensor (lbl) were carried out in the 1960-1980s
\cite{CTP1971} (with references therein). Using the general formalism developed in these papers a series
of applications was build for the cases when one, two, three and four photons were considered
as real ones.

Among them are the elastic scattering of photon on photon, elastic scattering of real
photon on Coulomb field of nuclei \cite{AB81}, splitting of photon to two real photons
in the Coulomb field \cite{197}, annihilation of electron and positron to three gluon jets
\cite{Baier}, neutral pion decay to four real photons \cite{164}, and the contribution to the
anomalous magnetic moment of muon \cite{143}.
The intermediate states with hadrons including the Higgs boson were
considered in a series of papers \cite{Jikia}. Recently, much attention was paid to
the problem of calculation of the hadronic contribution to the anomalous magnetic moment of the muon \cite{Bartosh},
where the intermediate states with the light scalar and pseudoscalar mesons were shown to be relevant.

In these works definite improvements results \cite{KN1950,CTP1971} was used. One of important applications concern
the lbl tensor forward scattering kinematics off mass shell photons. The improvement of (2) becomes too
complicate problem. Here we develop the independent approach using the  light-cone basis.

The special form of the lbl tensor (forward scattering of the virtual photons) was used in the description of
the peripheral collisions of hadrons \cite{Cheng}, investigating, in particular, the unitarity problems in QED.

A similar form of the lbl tensor was used in the problem of calculation of the Gell-Mann $\beta$ function \cite{Rafael}.

We mention as well the problems with construction and checking the Monte - Carlo generators to describe inelastic
processes in high energy hadrons (lepton-hadron) collisions.
These reasons are the motivations of our paper.

One of possible applications creation of lepton or pion pairs by two virtual photon of any possible polarization
mechanism at collision of any (charged) particles. Creation of several pairs by the some mechanism.
Constructing of the relevant Monte-Carlo generators \cite{Cheng}.
Another one-an alternative approach to study of polarization of vacuum operator in gauge theory in two loop
approximation $\Pi(q)$. In this case we have lbl tensor depending on two momenta - one of external boson,
$q$ and other, $k$-the momentum of the internal vector boson. A further integration $k$ is implied \cite{Rafael}.
We mention as well the problem of calculation of the photon (virtual) impact factor \cite{Cheng}.

For these purposes, the explicit form of the lbl tensor for the forward scattering kinematics can be useful.
It is the motivation of this paper.
A rather compact expression for the $s$- channel discontinuity of the lbl tensor written in the  explicit gauge
invariant form and obeying the Bose-symmetry is presented below (see (\ref{16})).
%It is the aim of our paper-to calculate the $s$ channel
The $s$ - channel discontinuity of the $lbl$ tensor
%with fermions or scalar particles intermediate state,
is associated with the scattering process
\ba
\gamma^*(k,\mu)+\gamma^*(q,\nu) \to \gamma^*(k,\mu_1)+\gamma^*(q,\nu_1),
\ea
namely,
\ba
\Delta_sL_{\mu\nu;\mu_1\nu_1}(k,q;k,q)=(4\pi\alpha)^2\int T_{\mu\nu}(k,q;q_+,q_-)(T_{\mu_1\nu_1}(k,q;q_+,q_-))^*d\Gamma_2,
\ea
with the phase volume of the intermediate state with two on mass shell charged particles
\ba
\int d\Gamma_2=\frac{1}{(2\pi)^2} \int d^4q_-d^4q_+\delta(q_-^2-m^2)\delta(q_+^2-m^2)\delta^4(k+q-q_+-q_-)= \nn \\
\frac{1}{(2\pi)^2}\frac{d^3q_+}{2E_+}\frac{d^3q_-}{2E_-}\delta^4(k+q-q_+-q_-).
\ea
Write down the phase volume in the form (we will work in the center of mass reference frame $\vec{k}+\vec{q}=0$):
\ba
\int d\Gamma_2=\frac{\beta}{8\pi}\frac{1}{2\pi}\int\limits_0^{2\pi}d\phi\frac{1}{2}\int\limits_{-1}^1d c,
\ea
with the 4-vector $q_-$ component defined as $q_-=(\sqrt{s}/2)(1,\beta\vec{n})$, $\vec{n}=(c, \sin\theta \cos\varphi, \sin\theta \sin\varphi)$ and
\ga{
s=(k+q)^2=(q_++q_-)^2;
\qquad
\beta=\sqrt{1-\frac{4m^2}{s}}, c=\cos\theta.
}
Here $\theta$ is the polar angle between the directions $\vec{k},\vec{q}_-$ and $\phi$ is the azimuthal angle which define the direction of the
3-vector $\vec{q}_-$ in the frames with the $z$ axis directed along $\vec{k}$.
The integration over the phase volume is essentially the angular averaging:
\eq{
\Delta_sL^{\mu\nu;\mu_1\nu_1}(k,q;k,q)&=(4\pi\alpha)^2\frac{\beta}{8\pi} l^{\mu\nu\mu_1\nu_1}, \nn \\
l^{\mu\nu\mu_1\nu_1}&=<T^{\mu\nu}(k,q;q_+,q_-)(T^{\mu_1\nu_1}(k,q;q_+,q_-))^*>, \nn \\
<F>&=\frac{1}{2\pi}\int\limits_0^{2\pi}d\phi\frac{1}{2}\int\limits_{-1}^1d c F(c,\phi).
}
\section{Light-like basis. General form of the zero-angle scattering tensor}
It is convenient to introduce two linear combinations of the photon 4-momenta $k=(k_1, k_2, k_3, k_4)=(k_0,|\vec{k}|,0,0); \,\,\,q=(q_0,-|\vec{k}|,0,0)$
\ga{
\chi =\frac{k+q}{\sqrt{s}}=(1,0,0,0), \qquad
r=\frac{q_0k-k_0q}{|\vec{k}|\sqrt{s}}=(0,1,0,0) \nn \\
\chi^2=1, \qquad r^2=-1, \qquad (\chi r)=0
}
and the "transversal" metric tensor
\ba
g^\bot_{\mu\nu}=\delta_{\mu,3}\delta_{\nu,3}+\delta_{\mu,4}\delta_{\nu,4}.
\ea
It has only one component transversal to 4-vectors $k,q$
\ba
g^\bot_{\mu\nu}k_\mu=g^\bot_{\mu\nu}q_\mu=0.
\ea
The transversal tensor can be written as
\ba
g^\bot_{\mu\nu}=-g_{\mu\nu}+\chi_\mu \chi_\nu-r_\mu r_\nu.
\ea

The photon momenta can be written in terms of the 4-vectors $\chi,r$ as
\ba
k = k_0\chi +p r, \qquad q = q_0\chi - p r, \qquad p=\frac{1}{2}\sqrt{\frac{\Lambda}{s}}.
\ea
The explicit form of $k_0,q_0,|\vec{k}|=|\vec{q}|, \Lambda=\Lambda(s,k^2,q^2)$ is given in Appendix A.

Besides, we introduce two vectors orthogonal to $k,q$:
\ba
\bar{k} =p\chi+k_0 r; \qquad \bar{q} =p\chi-q_0 r; \qquad (k\bar{k})=0; \qquad (q\bar{q})=0.
\ea

The $l^{\mu\nu\mu_1\nu_1}$ tensor can be written in terms of $g^\bot$ and $\bar{k},\bar{q}$ which
are explicit gauge-invariant quantities:
\ba
l^i_{\mu\nu\mu_1\nu_1}=a^i_F F+a^i_G G +a^i_H H+a^i_a a+a^i_b b+a^i_c c+a^i_d d+a^i_e e+a^i_f f +a^i_h\bar{k}_\mu\bar{k}_{\mu_1}\bar{q}_\nu\bar{q}_{\nu_1}, \qquad i=\pi,\mu,
\label{16}
\ea
with the $c$-number coefficients $a^i_j$ given below and the tensor structures
\eq{
F&=g^\bot_{\mu\mu_1}g^\bot_{\nu\nu_1}; &G&=g^\bot_{\mu\nu_1}g^\bot_{\nu\mu_1}; &H&=g^\bot_{\mu\nu}g^\bot_{\mu_1\nu_1}; \nn \\
a&=g^\bot_{\mu\mu_1}\bar{q}_\nu\bar{q}_{\nu_1}; &b&=g^\bot_{\mu\nu}\bar{q}_{\nu_1}\bar{k}_{\mu_1}; &c&=g^\bot_{\nu\mu_1}\bar{q}_{\nu_1}\bar{k}_{\mu}; \nn \\
d&=g^\bot_{\mu\nu_1}\bar{q}_\nu\bar{k}_{\mu_1}; &e&=g^\bot_{\nu_1\mu_1}\bar{q}_\nu\bar{k}_{\mu}; &f&=g^\bot_{\nu\nu_1}\bar{k}_\mu\bar{k}_{\mu_1}.
}
\section{Results}
In the case of pion pair production we obtain (details in Appendices A,B,C):
\eq{
a_H^\pi&=<4+16\lambda\frac{1}{d}\beta_1+16\frac{1}{d^2}\lambda^2\beta_2>; \nn \\
a_G^\pi=a_F^\pi&=16<\frac{1}{d^2}\lambda^2\beta_2>; \nn \\
p^2a_b^\pi=p^2a_e^\pi&=<4[\frac{\lambda}{d}-4\frac{\lambda^2}{d^2}]\beta_1-4+2\frac{\lambda}{d}[-5+\rho]>; \nn \\
p^2a_d^\pi=p^2a_c^\pi&=<4\lambda^2(\frac{1}{d}-\frac{1}{d^2})(\rho-1)\beta_1>; \nn \\
p^2a_a^\pi&=<4\lambda^2\eta^2(\frac{1}{d}-\frac{1}{d^2})\beta_1>; \nn \\
p^2a_f^\pi&=<4\lambda^2\sigma^2(\frac{1}{d}-\frac{1}{d^2})\beta_1>; \nn \\
p^4a_h^\pi&=<4+4(1-\rho)\frac{\lambda}{d}+4[16+(1-\rho)^2]\frac{\lambda^2}{d^2}>.
}

The coefficients of the tensor for the lepton pair in the intermediate state are (index $\mu$ corresponds to $\mu$ - meson)
\ba
p^2a_a^\mu=\frac{8\lambda^2}{d^2}[(1-d)\epsilon^2+[\frac{1}{4}d(\delta-\epsilon-1)^2-\lambda^2(\delta+\epsilon-1)^2\epsilon]\beta_1]; \nn \\
p^2a_b^\mu=-\frac{4(1-d)}{d^2}[\frac{1}{2}d-2\lambda^2\epsilon\delta+\lambda^2(\delta+\epsilon-1)^2\beta_1]; \nn \\
p^2a_c^\mu=-\frac{4\lambda^2}{d^2}[2(1-d)\epsilon\delta+[\frac{1}{2}(2-d)(\delta-\epsilon-1)^2-(1-d)(\delta+\epsilon-1)-2\epsilon(2-d)]\beta_1]; \nn \\
a_F^\mu=\frac{8\lambda^2(\epsilon+\delta-1)^2}{d^2}[\frac{1}{4}d-\epsilon\delta\lambda^2+(\delta+\epsilon)\lambda^2\beta_1-\lambda^2\beta_2]; \nn \\
a_G^\mu=-\frac{8\lambda^2(\epsilon+\delta-1)^2}{d^2}[\frac{1}{4}d-\epsilon\delta\lambda^2+(\delta+\epsilon-1)\lambda^2\beta_1+\lambda^2\beta_2]; \nn \\
a_H^\mu=\frac{8(1-d)}{d^2}[[\frac{1}{4}d-\epsilon\delta\lambda^2+(\delta+\epsilon-1)\lambda^2\beta_1](1-d)-\lambda^4(\epsilon+\delta-1)^2\beta_2]; \nn \\
p^4a_h^\mu=\frac{8(1-d)}{d^2}[\frac{1}{4}d-\epsilon\delta\lambda^2].
\ea
Moreover,
\ba
a_f^\mu=a_a^\mu(\epsilon \leftrightarrow \delta); \,\,\,p^2a_e^\mu=a_b^\mu;\,\,\,a_d^\mu=a_c^\mu.
\ea
Here we use the notation for scalar coefficients
\ba
\epsilon=\frac{k^2}{s}; \,\,\,\delta=\frac{q^2}{s};\,\,\,\sigma=\frac{q_0}{\sqrt{s}};\,\,\,\eta=\frac{k_0}{\sqrt{s}}; \nn \\
\lambda=\frac{1}{\epsilon+\delta-1}; \,\,\,\rho=\frac{\Lambda}{s^2}, \,\,\,\beta_1=\frac{1}{2}\beta^2(1-c^2); \,\,\, \beta_2=\frac{1}{8}\beta^4(1-c^2)^2.
\ea
The relevant integrals needed to perform the integration over the polar angle are given in Appendix C.

\section{conclusion. The case of real photons}

In the case of both photons on the mass shell the 4-vectors $\bar{k},\bar{q}$ become the light-like ones. Using
\ba
p=\frac{\sqrt{s}}{2}; \,\,\,\rho=1,\,\,\,\lambda=-1; \,\,\,\eta=\sigma=\frac{1}{2}, \,\,\,\epsilon=\delta=0,
\ea
we obtain
\eq{
a_H^\pi=<4-16\frac{1}{d}\beta_1+16\frac{1}{d^2}\beta_2>; \nn \\
a_G^\pi=a_F^\pi=16<\frac{1}{d^2}\beta_2>; \nn \\
\frac{s}{4}a_b^\pi=\frac{s}{4}a_e^\pi=<-4\beta_1[\frac{1}{d}+4\frac{1}{d^2}]-4+8\frac{1}{d}>; \nn \\
\frac{s}{4}a_d^\pi=p^2a_c^\pi=0; \nn \\
\frac{s}{4}a_a^\pi=\frac{s}{4}a_f^\pi=<\beta_1(\frac{1}{d}-\frac{1}{d^2})>; \nn \\
\frac{s^2}{16}a_h^\pi=4<1+16\frac{1}{d^2}>.
}
and
\ba
a_F^\mu=<\frac{8}{d^2}[\frac{1}{4}d-\beta_2>; \nn \\
a_G^\mu=<-\frac{8}{d^2}[\frac{1}{4}d-\beta_1+\beta_2>; \nn \\
a_H^\mu=<\frac{8}{d^2}[(1-d)(\frac{1}{4}d-\beta_1)-\beta_2>; \nn \\
a_d^\mu=a_a^\mu=-a_c^\mu=-a_d^\mu=\frac{2}{d}\beta_1; \nn \\
a_b^\mu=a_e^\mu=\frac{4}{d^2}(1-d)(\frac{1}{2}d-\beta_1), \nn \\
d=1-\beta^2c^2.
\ea

An important test is the correspondence with the cross sections of pair production in real photon collisions.
Really,
\ba
\frac{d\sigma^{\gamma\gamma\to a\bar{a}}}{d c}=\frac{\alpha^2}{2(s-4m^2)}g_{\mu\mu_1}g_{\nu\nu_1}l_{\mu\nu\mu_1\nu_1}.
\ea
We obtain for the cross section of a pair of charged scalar particle production \cite{AB81}
\ba
\frac{d\sigma}{d c}=\frac{\alpha^2}{2s}\beta[1-\frac{8m^2}{s}\frac{1}{d}+\frac{32m^2}{s^2}\frac{1}{d^2}],
\ea
and for a fermion pair production
\ba
\frac{d\sigma}{d c}=\frac{2\alpha^2\beta}{s}[\frac{1+\beta^2c^2}{d}+\frac{8m^2}{s d}-\frac{32m^4}{s^2d^2}].
\ea
Constructing the combination $g^{\mu\mu_1}g^{\nu\nu_1}l_{\mu\nu\mu_1\nu_1}=4a_F+2(a_G+a_H)$ one
can be convinced that these tests are fulfilled.

We note in conclusion that the tensors (\ref{16}) being converted with the light-like vectors $\frac{1}{\sqrt {2}}(\chi \pm r)_{\mu} = V_{\mu}^{\pm}: l_{\mu \mu_1 \nu \nu_1}^i V_{\mu}^+ V_{\mu_1}^+ V_{\nu}^- V_{\nu_1}^-$ are the kernels of
integral equations for zero angles scattering amplitudes, investigated in \cite{Cheng}.

The general form of lbl tensor (15) for the forward scattering kinematics is our main result. It satisfy all the
symmetry requirements and have rather compact form. The special realization for definite intermediate states are
considered.

\section{acknowledgements}

One of us (E.K.) acknowledges the support of RFBR, grants 10-02-01295a; 11-02-00112.

\appendix

%%%%%%%%%%%%%%%%%%%%%%%%%%%%%%%%%%%%%%%%%%%%%%%%%%%%%%

\section{The charged pi-meson pair intermediate state. Kinematics}

The matrix element of the subprocess has the form
\ba
M_{scal}^{\gamma^* \gamma^* \to q^- q^+} =  4\pi\alpha T_{\mu\nu},
\ea
with
\ba
T_{\mu\nu} = \frac{(2q_- -k)_{\mu}(-2q_+ +q)_{\nu}}{d_-} + \frac{(k-2q_+)_{\mu}(2q_- -q)_{\nu}}{d_+} -2g_{\mu\nu}, \nn \\
d_- = (q_- -k)^2 -m^2 = k^2-2q_- k; \,\,\,d_+ = (-q_+ +k)^2 -m^2 =k^2-2q_-q.
\ea
One can check the fulfilment of the gauge-invariance condition $T_{\mu\nu}k_\mu=T_{\mu\nu}q_\nu=0$.

It seems to be convenient to rewrite the Dirac tensor $T_{\mu\nu}$ as
\ba
T_{\mu\nu} = Aq_{{-}{\mu}}q_{{-}{\nu}} - 2q_{{-}{\mu}}C_{\nu} -2q_{{-}{\nu}}D_{\mu} +r_{\mu\nu} +2g_{\mu\nu}^{\bot}; \nn \\
A = \frac{1}{d_-} +\frac{1}{d_+}; \,\,\,\,C_{\nu} = \biggl(\frac{2k +q}{d_-} +\frac{q}{d_+}\biggr)_{\nu};
\,\,\,D_{\mu} =\biggl(\frac{k}{d_-} +\frac{2q +k}{d_+}\biggr)_{\mu};  \nn \\
r_{\mu\nu} = \frac{1}{d_-}k_{\mu}(2k+q)_{\nu} + \frac{1}{d_+}q_{\nu}(2q+k)_{\mu} -2\chi_{\mu}\chi_{\nu} +2r_{\mu}r_{\nu}; \nn \\
g_{\mu\nu}^{\bot} = -g_{\mu\nu} +\chi_{\mu}\chi_{\nu} -r_{\mu}r_{\nu}.
\ea

Using the energy-momentum conservation law
\ba
k + q = q_+ + q_- \nn
\ea
we find
\ba
k_0 = \frac{s+k^2 - q^2}{2\sqrt{s}}, \,\,\,q_0 =\frac{s + q^2 - k^2}{2\sqrt{s}}; \nn \\
\vec{k}^2=\frac{\Lambda}{4s}, \Lambda=\Lambda(s,k^2,q^2)=s^2+(k^2)^2+(q^2)^2-2sk^2-2sq^2-2k^2q^2.
\ea

Below we will express the lbl tensor in terms of $g^\bot$ and $\bar{k},\bar{q}$ which
are explicit gauge-invariant quantities.

So we have (see (9)):
\ba
\Delta_sL_{\mu\nu;\mu_1\nu_1}(k,q;k,q)=\frac{\beta}{8\pi}(4\pi\alpha)^2<{T_{\mu\nu}(T_{\mu_1\nu_1})^*}>.
\ea

\section{The charged lepton pair intermediate state}

The matrix element of the subprocess has the form
\ba
M_{q_+q_-}^{\gamma^* \gamma^* \to q^- q^+} =  4\pi\alpha T_{{\mu\nu}({q_+q_-})},
\ea
with
\ba
T_{{\mu\nu}({q_+q_-})} = \bar{u}(q_-)[\gamma_\mu\frac{\hat{q}_--\hat{k}+m}{d_-}\gamma_\nu+\gamma_\nu\frac{-\hat{q}_++\hat{k}+m}{d_+}\gamma_\mu] v(q_+), \,\,\,\,\hat{a} = a_{\mu}\gamma_{\mu}. \nn
\ea
Using the on mass shell conditions of leptons it can be written as
\ba
T_{{\mu\nu}({q_+q_-})} = \bar{u}(q_-)[Q_\mu \gamma_\nu-\frac{1}{d_-}\gamma_\mu\hat{k}\gamma_\nu+\frac{1}{d_+}\gamma_\nu\hat{k}\gamma_\mu] v(q_+), \nn \\
Q_\mu=2(\frac{q_-}{d_-}-\frac{q_+}{d_+})_\mu.
\ea
Rather tedious calculations lead to the result given above.

\section{details of calculations}

The expression for the denominators are
\ba
d_- =\frac{k^2 + q^2 -s}{2}(1+b c),\,\,\,d_+=\frac{k^2 + q^2 -s}{2}(1-b c),\,\,\,
 b = \beta\frac{\sqrt{\Lambda}}{k^2 + q^2 -s},\,\,c=\cos\theta,\,\,\theta=\widehat{ \vec{q}_-
 \vec{k}}.
\ea
The product of two Dirac tensors for a pion pair has the form (notation in (A3)):
\ba
T_{\mu\nu}T_{\mu_1\nu_1} = 16A^2(q_-q_-q_-q_-)_{\mu\mu_1\nu\nu_1} - 8AC_{\nu_1}(q_-q_-q_-)_{\mu\nu\mu_1} -
8AC_{\nu}(q_-q_-q_-)_{\mu\mu_1\nu_1} - 8AD_{\mu_1}(q_-q_-q_-)_{\mu\nu\nu_1} - \nn \\
8AD_{\mu}(q_-q_-q_-)_{\nu\mu_1\nu_1}+4Ar_{\mu_1\nu_1}(q_-q_-)_{\mu\nu}+4Ar_{\mu\nu}(q_-q_-)_{\mu_1\nu_1}+  \nn \\ 4C_{\nu}C_{\nu_1}(q_-q_-)_{\mu\mu_1}+4D_{\mu}D_{\mu_1}(q_-q_-)_{\nu\nu_1} + 4C_{\nu}D_{\mu_1}(q_-q_-)_{\mu\nu_1} + 4C_{\nu_1}D_{\mu}(q_-q_-)_{\nu\mu_1} - \nn \\
2C_{\nu}r_{\mu_1\nu_1}q_{-{\mu}}-2C_{\nu_1}r_{\mu\nu}(q_-)_{\mu_1} -2 D_{\mu}r_{\mu_1\nu_1}q_{-{\nu}} -
2D_{\mu_1}r_{\mu\nu}q_{-{\nu_1}} +r_{\mu\nu}r_{\mu_1\nu_1} + 4g_{{\bot}{\mu\nu}}q_{{\bot}{\mu_1\nu_1}}+ \nn \\
2g_{{\bot}{\mu\nu}}[4A(q_-q_-)_{\mu_1\nu_1} -2C_{\nu_1}q_{-{\mu_1}}-2D_{\mu_1}q_{-{\nu_1}}+r_{\mu_1\nu_1}]+
2g_{{\bot}{\mu_1\nu_1}}[4A(q_-q_-)_{\mu\nu}-2q_{-{\mu}}C_{\nu} -2q_{-{\nu}}D_{\mu}+r_{\mu\nu}].
\ea
The averaging of the relevant 4-vector product gives
\eq{
<q_{-{\mu}}>&=\frac{\sqrt{s}}{2}<(\chi +\beta cr)_{\mu}>; \nn\\
<q_{-{\mu}}q_{-{\nu}}>&=\frac{s}{4}<[\chi^2+\beta c(\chi r) +\beta^2(c^2r^2 +\frac{1}{2}(1-c^2)g_{\bot})]_{\mu\nu}>; \nn \\
<q_{-{\mu}}q_{-{\nu}}q_{-{\lambda}}>&=\biggl(\frac{\sqrt{s}}{2}\biggr)^3
<\{\chi^3 +(\chi^2 r)\beta c +\beta^2[(\chi r^2) c^2 +\frac{1}{2} (1-c^2)(\chi g_{\bot})] +
\beta^3[c^3r^3 +\frac{1}{2}c (1-c^2)(rg_{\bot})]\}_{\mu\nu\lambda}>; \nn \\
<q_{-{\mu}}q_{-{\nu}}q_{-{\lambda}}q_{-{\sigma}}> &= \biggl(\frac{s}{4}\biggr)^2 <\{\chi^4 +\beta c (\chi^3 r) + \nn \\
&\beta^2[c^2 (\chi^2 r^2) +\frac{1}{2}(1-c^2)(\chi^2g_{\bot})] +\beta^3[c^3(\chi r^3) +
\frac{1}{2} c (1-c^2)(\chi r g_{\bot})] + \nn \\
&\beta^4[c^4 r^4 +\frac{1}{2} (c^2 (1-c^2))(r^2 g_{\bot}) +\frac{1}{8} (1-c^2)^2(g_{\bot}g_{\bot})]\}_{\mu\nu\mu_1\nu_1}>.
}
Here we imply
\ba
(a b)_{\mu\nu}=a_\mu b_\nu+a_\nu b_\mu; \,\,(a^n)_{\mu_1...\mu_n}=a_{\mu_1}...a_{\mu_n}; \,\,
(a^2 b)_{\mu_1\mu_2\mu_3}=a_{\mu_1}a_{\mu_2}b_{\mu_3}+...+b_{\mu_1}a_{\mu_2}a_{\mu_3}; \nn \\
(a g_\bot)_{\mu_1\mu_2\mu_3}=a_{\mu_1}g_{\bot {\mu_2\mu_3}}+a_{\mu_2}g_{\bot {\mu_1\mu_3}}+a_{\mu_3}g_{\bot {\mu_2\mu_1}}; \,\,
(a^2 b^2)_{\mu_1\mu_2\mu_3\mu_4}=a_{\mu_1}a_{\mu_2}b_{\mu_3}b_{\mu_4}+...+b_{\mu_1}b_{\mu_2}a_{\mu_3}a_{\mu_4}; \nn \\
(a^2g_\bot)_{\mu_1\mu_2\mu_3\mu_4}=a_{\mu_1}a_{\mu_2}g_{\bot {\mu_2\mu_3}}+a_{\mu_1}g_{\bot {\mu_1\mu_3}}a_{\mu_4}+...+g_{\bot {\mu_1\mu_2}}a_{\mu_3}a_{\mu_4}; \nn \\
(g_\bot g_\bot)_{\mu_1...\mu_4}=g_{\bot {\mu_1\mu_2}} g_{\bot {\mu_3\mu_4}}+g_{\bot {\mu_1\mu_3}} g_{\bot {\mu_2\mu_4}}+g_{\bot {\mu_1\mu_4}} g_{\bot {\mu_2\mu_3}}.
\ea

The relevant integrals needed to perform the integration over the polar angle are ($d=1-b^2c^2$):
\ba
\varphi_1=<\frac{1-c^2}{d}>= \frac{1}{b^2}\biggl(1-\frac{1-b^2}{2b} L\biggr); \,\,\,\,
\varphi_2=<\frac{(1-c^2)^2}{d^2}= \frac{1}{2b^4}\biggl[3-b^2-\frac{1-b^2}{2b}(3+b^2)L\biggr]; \nn \\
\varphi_3=<\frac{1}{d}>= \frac{1}{b}L;
\varphi_4= <\frac{c^2(1-c^2)}{d^2}>=\frac{1}{b^4}[-\frac{3}{2}+\frac{1}{4}(7-3b^2)\frac{1}{b}L\biggr]; \nn \\
\varphi_5=<\frac{1}{d^2}>= \frac{1}{1-b^2} + \frac{1}{2b}L; \,\,\,\,
\varphi_6=<\frac{c^2}{d^2}>=\frac{1}{b^2}[\frac{1}{1-b^2}-\frac{1}{2b}L];\,\,\,\,
L = \ln\frac{1+b}{1-b}.
\ea
For the charged $\pi$-meson case we have
\ba
a_H^\pi = 4 +4\beta^2 \lambda \varphi_1 +\beta^4 \lambda^2 \frac{1}{2}\varphi_2; \nn \\
a_G^\pi =a_F^\pi = \frac{1}{2}\beta^4 \lambda^2 \varphi_2; \nn \\
p^2a_b^\pi=p^2a_e^\pi= -4+\beta^2 \lambda (\varphi_1 -2 \varphi_2 \lambda) +\varphi_3 \lambda (-5 + \rho);  \nn \\
p^2a_c^\pi=p^2a_d^\pi= \frac{1}{2}\beta^2 \lambda^2 (\rho -1)b^2 \varphi_4;  \nn \\
p^2a_a^\pi= \frac{1}{2}\beta^2 \eta^2\lambda^2 b^2 \varphi_4;  \nn \\
p^2a_f^\pi= \frac{1}{2}\beta^2\sigma^2\lambda^2 b^2 \varphi_4;  \nn \\
p^4a_h^\pi = 4+2\lambda(1-\rho)\varphi_3 +(16 +(1-\rho)^2)\lambda^2 \varphi_5.
\ea

%
%\begin{figure}
%\centering
%{\includegraphics[width=0.8\textwidth]{Fig1.eps}}
% \caption{\label{Fig:1ab}
%One vector meson production in ions collision.
%}
%\end{figure}
%
%
%\begin{figure}
%\centering
%{\includegraphics[width=0.8\textwidth]{Fig2.eps}}
% \caption{\label{Fig:1ab}
%Two vector meson production: a) intermediate vector meson state; b) intermediate
%two gluon state.
%}
%\end{figure}
%


\begin{thebibliography}{10}
\bibitem{KN1950}
R.~Karplus and M.~Neuman,
Phus.REv. 80,(1950),380.
\bibitem{CTP1971}
V.~Costantini, B.~De-Tollis and G.~Pistoni,
Nuovo Cimento, 2A(1971),733.
\bibitem{AB81}
A.~I.~Akhiezer and V.~B.~Berestetskij,
Quantum Electrodynamics, Chapter 5, Nauka, Moscow, 1981.
\bibitem{197}
V.~N.~Baier, V.~S.~Fadin, V.~M.~Katkov and E.~A.~Kuraev,
Phys. Lett. B49 (1974)385; \\
V.~N.~Baier, A.~I.~Milshtein, R.~Zh.~Shaisultanov,
Phys. Rev. Lett. 77 (1996)1691; \\
R.~N.~Li, A.~I.~Milshtein, V.~M.~Strakhovenko,
J.Exp.Theor.Phys. 85(1997)1649; \\
E.~Kuraev and S.~Sannikov,
ZETP,44(1963),1015.
\bibitem{Baier}
V.~N.~Baier, E.~A.~Kuraev, V.~S.~Fadin,
Yad. Fiz. 31 (1980)700.
\bibitem{164}
E.~A.~Kuraev, Z.~K.~Silagadze, A.~A.~Cheshel and A.~Schiller,
Sov. J. Nucl. Phys. 50 (1989)264; Yad. Fiz. 50, (1989)422.
\bibitem{143}
E.~L.~Bratkovskaya, E.~A.~Kuraev, Z.~K.~Silagadze,
Phys. Lett. B359 (1995)217.
\bibitem{Jikia}
R.~Belusevic and G.~Jikia,
Phys. Rev. D70 (2004)073017.
\bibitem{Bartosh}
E.~Bartos, S.~Dubnicka, A.~Z.~Dubnickova, E.~A.~Kuraev, E.~Zemlyanaya,
hep-ph/0305051.
\bibitem{Cheng}
H.~Cheng and T.~T.~Wu,
Phys. Rev.182(1969)1852, 1868, 1873, 1899; \\
G.~V.~Frolov, V.~N.~Gribov, L.~N.~Lipatov,
Phys.Lett. 31B (1970)34.
\bibitem{Rafael}
B.~Lautrup and E. de Rafael,
Nucl. Phys. B70 (1974)317; \\
E. de Rafael and J.~N.~Rosner,
Annals Physics 82 (1974)369.

\end{thebibliography}
\end{document}